\documentclass[12pt]{article}
\usepackage{graphicx}
\usepackage{cite}

\graphicspath{{figs/}}

\usepackage{amsmath,amssymb}

\usepackage{url,comment}
\usepackage{graphicx, graphics, hyperref, amsmath, amssymb, slashed, xcolor, bbm,amsthm, array}
 \usepackage{subfigure}
\usepackage{rotating}
\usepackage{afterpage}
\newcommand{\nc}{\newcommand}
\nc{\beq}{\begin{equation}}
\nc{\eeq}{\end{equation}}
\nc{\barray}{\begin{eqnarray}}
\nc{\earray}{\end{eqnarray}}
\nc{\barrayn}{\begin{eqnarray*}}
\nc{\earrayn}{\end{eqnarray*}}
\nc{\bcenter}{\begin{center}}
\nc{\ecenter}{\end{center}}
\nc{\mc}{\mathcal}
\nc{\er}[1]{(\ref{eq:#1})}
\nc{\onehalf}{\frac{1}{2}} 
\nc{\partialbar}{\bar{\partial}}
\nc{\psit}{\widetilde{\psi}}
\nc{\Tr}{\mbox{Tr}}
\nc{\hc}{\mbox{H.c.}}
\nc{\ev}{\;\mathrm{eV}}
\nc{\mev}{\;\mathrm{MeV}}
\nc{\gev}{\;\mathrm{GeV}}
\nc{\tev}{\;\mathrm{TeV}}

\def\chii0{\chi_i^0}
\def\chij0{\chi_j^0}

\newcommand{\gsim}{\lower.7ex\hbox{$\;\stackrel{\textstyle>}{\sim}\;$}}% This is just \gtrsim
\newcommand{\lsim}{\lower.7ex\hbox{$\;\stackrel{\textstyle<}{\sim}\;$}}%This is just \lesssim
\nc{\ttbar}{t\bar t}

\newcommand{\fref}[1]{Fig.~\ref{f.#1}}
\newcommand{\eref}[1]{Eq.~(\ref{e.#1})}

\newcommand{\sref}[1]{Section~\ref{s.#1}}

\newcommand{\cref}[1]{Chapter~\ref{c.#1}}

\newcommand\pubnumber{SLAC--PUB--16981}
\newcommand\pubdate{May 2017}

\def\SLAC{SLAC,
    Stanford University, Menlo Park, California 94025 USA}
\def\doeack{\footnote{Work supported by the US Department of Energy,
                     contract DE--AC02--76SF00515.}}

\def\Title#1{\begin{center} {\Large #1 } \end{center}}
\def\Author#1{\begin{center}{ \sc #1} \end{center}}
\def\Address#1{\begin{center}{ \it #1} \end{center}}
\def\andauth{\begin{center}{and} \end{center}}
\def\submit#1{\begin{center}Submitted to {\sl #1} \end{center}}
\newcommand\pubblock{\rightline{\begin{tabular}{l} \pubnumber\\
         \pubdate \end{tabular}}}
\newenvironment{Abstract}{\begin{quotation} \begin{center}
                       ABSTRACT
     \end{center}\bigskip  }{\end{quotation}}

\def\submit#1{\begin{center}Submitted to {\sl #1} \end{center}}
\def\Acknowledgements{\bigskip  \bigskip \begin{center} \begin{large}
             \bf ACKNOWLEDGEMENTS \end{large}\end{center}}
\def\beq{\begin{equation}}
\def\eeq#1{\label{#1}\end{equation}}
\def\eeqn{\end{equation}}

\newenvironment{Eqnarray}%
   {\arraycolsep 0.14em\begin{eqnarray}}{\end{eqnarray}}
\def\beqa{\begin{Eqnarray}}
\def\eeqa#1{\label{#1}\end{Eqnarray}}
\def\eeqan{\end{Eqnarray}}

\def\leqn#1{(\ref{#1})}

\let\bar=\overbar

\def\lsim{\mathrel{\raise.3ex\hbox{$<$\kern-.75em\lower1ex\hbox{$\sim$}}}}
\def\gsim{\mathrel{\raise.3ex\hbox{$>$\kern-.75em\lower1ex\hbox{$\sim$}}}}

\def\hc{{\mbox{\rm h.c.}}}

\def\del{\partial}
\def\Dslash{\not{\hbox{\kern-4pt $D$}}}
\def\dslash{\not{\hbox{\kern-2pt $\del$}}}

\def\msb{{\bar{\scriptsize M \kern -1pt S}}}

\def\drb{{\bar{\scriptsize D \kern -1pt R}}}

\makeatletter
\def\section{\@startsection{section}{0}{\z@}{5.5ex plus .5ex minus
 1.5ex}{2.3ex plus .2ex}{\large\bf}}
\def\subsection{\@startsection{subsection}{1}{\z@}{3.5ex plus .5ex minus
 1.5ex}{1.3ex plus .2ex}{\normalsize\bf}}
\def\subsubsection{\@startsection{subsubsection}{2}{\z@}{-3.5ex plus
-1ex minus  -.2ex}{2.3ex plus .2ex}{\normalsize\sl}}

\renewcommand{\@makecaption}[2]{%
   \vskip 10pt
   \setbox\@tempboxa\hbox{\small #1: #2}
   \ifdim \wd\@tempboxa >\hsize     % IF longer than one line:
       \small #1: #2\par          %   THEN set as ordinary paragraph.
     \else                        %   ELSE  center.
       \hbox to\hsize{\hfil\box\@tempboxa\hfil}
   \fi}

\makeatother

\begin{document}
\begin{titlepage}
\pubblock

\vfill
\Title{Analysis of Long Lived Particle Decays \\ \vskip 0.1in
with the MATHUSLA Detector}
\vfill
\Author{David Curtin\footnote{Work supported by the National Science Foundation, grant
no. NSF-PHY-1620074.} }
\Address{Maryland Center for Fundamental Physics, Dept. of Physics \\ 
    University of Maryland, College Park, MD 20742 USA }
\andauth
\Author{Michael E. Peskin\doeack}
\Address{\SLAC}
\vfill
\begin{Abstract}
The MATHUSLA detector is a simple large-volume  tracking
detector to be located on the surface above one of the general-purpose
experiments at the Large Hadron Collider.    This detector was
proposed in \cite{Chou:2016lxi} to detect exotic, neutral, long-lived
particles that might be produced in high-energy proton-proton
collisions.  In this paper, we consider the use of the limited
information that MATHUSLA would provide on the decay products of the
long-lived particle.   For the case in which the long-lived particle
is pair-produced in Higgs boson decays, we show that it is possible to
measure the mass of this particle and determine the dominant decay
mode with less than $100$ observed events.  We discuss the
ability of MATHUSLA to distinguish the production mode of the
long-lived particle and to determine its mass and spin in more general cases.
\end{Abstract}
\vfill
\submit{Physical Review D}
\vfill
\newpage
\tableofcontents
\end{titlepage}

\def\thefootnote{\fnsymbol{footnote}}
\setcounter{footnote}{0}

\section{Introduction}

Despite the successes of the Standard Model of particle physics, there
are strong motivations to believe in new fundamental interactions that
lie outside this model.   The Standard Model does not contain a
particle that could explain the dark matter of the universe. Its theory
of the Higgs boson and its symmetry-breaking potential is completely
{\it ad hoc}.  Many models have been proposed to generalize the
Standard Model, but there is no compelling experimental evidence
supporting any of these models.   It is therefore important to propose
additional windows through which to search for these new interactions.

A property shared by many models of supersymmetry
\cite{Arvanitaki:2012ps,ArkaniHamed:2012gw,Giudice:1998bp,Barbier:2004ez,Csaki:2013jza,Fan:2011yu}, 
neutral naturalness \cite{Burdman:2006tz,Cai:2008au,Chacko:2005pe}, 
dark matter
\cite{Baumgart:2009tn,Kaplan:2009ag,Chan:2011aa,Dienes:2011ja,Dienes:2012yz,Kim:2013ivd}, 
baryogenesis
\cite{Bouquet:1986mq,Campbell:1990fa,Cui:2012jh,Barry:2013nva,Cui:2014twa,Ipek:2016bpf}, 
neutrinos
\cite{Helo:2013esa,Antusch:2016vyf,Graesser:2007yj,Graesser:2007pc,Maiezza:2015lza,Batell:2016zod,Dev:2016vle,Dev:2017dui,Nemevsek:2016enw,Caputo:2017pit},
and 
Hidden Valleys
\cite{Strassler:2006im,Strassler:2006ri,Strassler:2006qa,Han:2007ae,Strassler:2008bv,Strassler:2008fv}
is that
 they contain long-lived particles (LLPs)  with
macroscopic decay lengths.   

Searches for LLPs have typically involved
the study of low-energy reactions, for example, using fixed target
experiments with electron, proton, or neutrino beams.  
One strategy has been to position a detector 
behind a beam dump, where it can observe decays of neutral particles
with weak interaction cross sections on matter. 
However, this approach to the search of LLPs is limited in mass scale. It is also limited because it
requires the LLP to have large enough coupling to quarks and leptons.

The Large Hadron Collider (LHC) offers new mechanisms for the
production of LLPs that are available only in high-energy collisions.
These include production through  $W$ boson fusion, through the decay
of heavy SM particles like the Higgs or $Z$, through
 the decay of new heavy parent particles such as squarks or gluinos,
 and through new, heavy scalar and vector bosons produced in the $s$-channel in
quark-antiquark or gluon-gluon collisions. 
The most interesting and most highly motivated of these mechanisms is the exotic decay of
the 125~GeV Higgs boson to a pair of LLPs \cite{Curtin:2013fra,Curtin:2014cca,Craig:2015pha,Curtin:2015fna}.   However, though the LHC might have large production rates
for LLPs, the ability of the LHC detectors to observe these particles
is limited.  As large as ATLAS and CMS are, the size of these
detectors is a constraint.  Furthermore, LLP events suffer from 
significant  backgrounds,
especially if the LLPs decay to hadrons \cite{Coccaro:2016lnz}.

The MATHUSLA detector was proposed in~\cite{Chou:2016lxi} to address this
problem \cite{Maki:1997ih}.
 MATHUSLA is a large-volume
detector on the surface above an LHC experiment. 
Essentially, it is an empty barn that provides a decay volume for LLPs, and, near its
roof, is equipped with charged particle tracking to detect an LLP decay. 
It is shown in ~\cite{Chou:2016lxi} that the limited instrumentation
proposed allows one to reject cosmic-ray and other backgrounds with very
high confidence.   This dedicated detector would  increase the
sensitivity to LLPs over the capabilities of the current central
detectors
 by several orders of magnitude.  The comparison to ATLAS is shown in 
 \fref{mathuslareach}.

Because of its large size and because --- as yet --- there is no evidence
for LLPs,
the MATHUSLA detector  must be built from relatively inexpensive components.   
The original concept for MATHUSLA in \cite{Chou:2016lxi} imagined an empty building offering 20 m of decay space and, above this, $\sim 5$ layers of Resistive Plate Chambers (RPCs), along with some plastic scintillator for additional timing and veto information.
This paper offered an explicit physics case for MATHUSLA, with
estimates of its performance
 in the search for LLPs produced in exotic Higgs decays as a
 well-motivated benchmark
 model\cite{whitepaper} 

From this description, it is not obvious that MATHUSLA has
any capability beyond the discovery of LLP events via the detection of
decay vertices originating in its decay volume.  However, we find
that,  
by applying  some simple arguments, it is possible to use the limited
information provided by MATHUSLA to learn a surprising amount.   
In this paper, we analyze the performance of MATHUSLA for the most
interesting and also most constrained situation---the decay of the
Higgs boson to a pair of LLPs, such that the LLP has a dominant 2-body
decay mode. In Section 2, we briefly review the design of MATHUSLA.
In Section 3, we show that, under the assumption of this production mode, 
it is possible to measure the mass of the LLP 
and to identify its most important decay modes, using only 
the information provided by MATHUSLA, with as few as $ 30 - 100$
observed 
decays.  In Section 4, we show how the production by Higgs decay
may be distinguished from other hypotheses, and we discuss the
generalization of this analysis to other LLP production modes.

\begin{figure}
\begin{center}
\includegraphics[width=12cm]{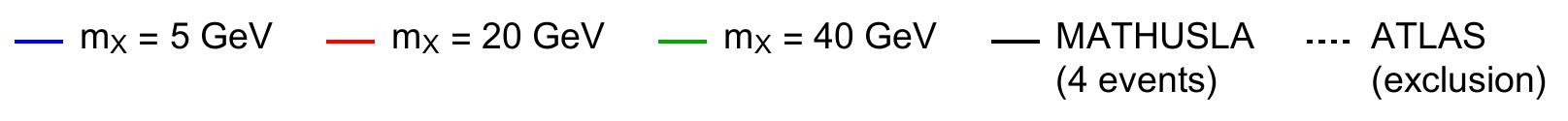}
\\
\includegraphics[width=12cm]{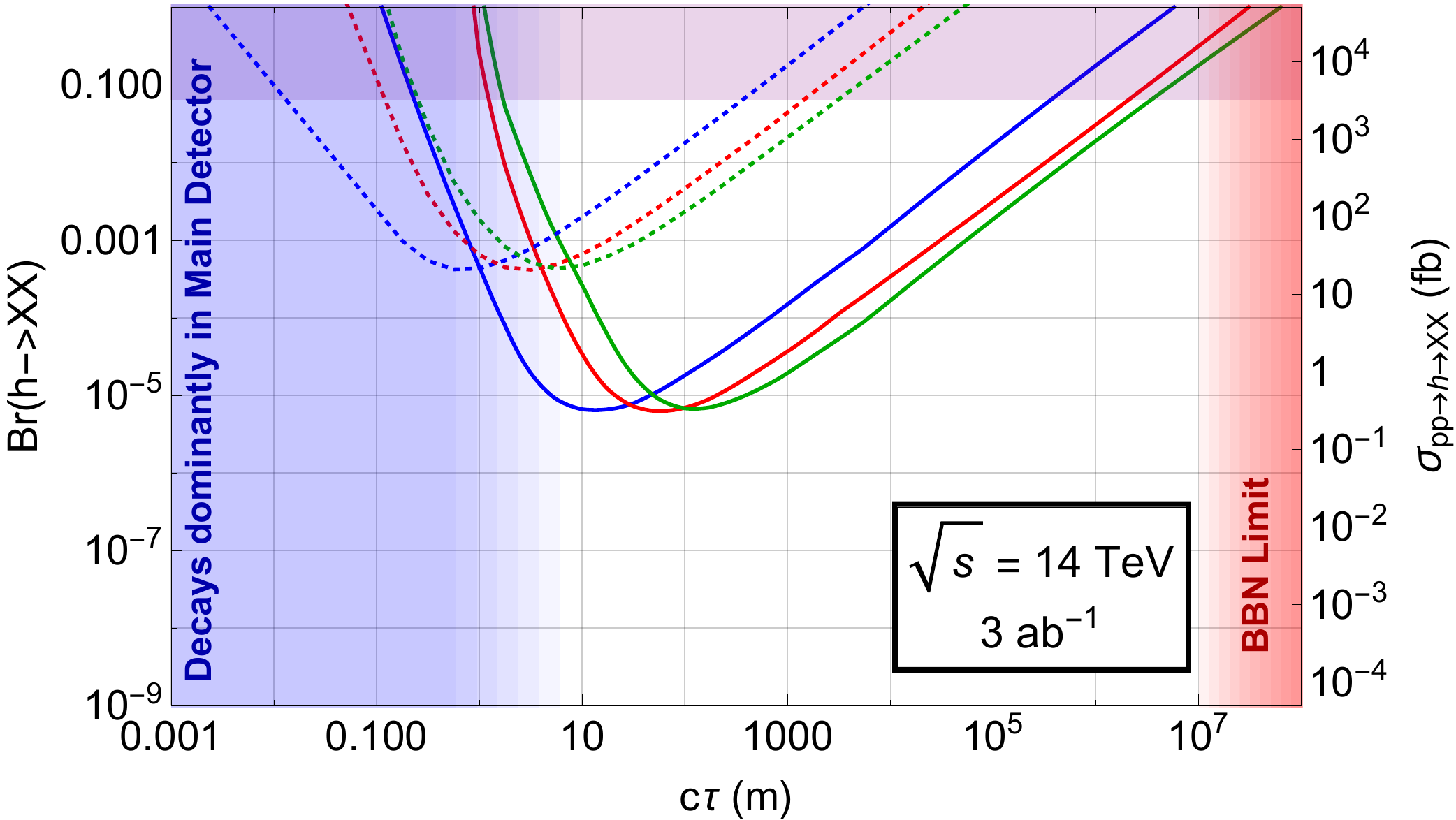}
\end{center}
\caption{
Exclusion reach of MATHUSLA, corresponding to 4 expected decays
in the detector (solid curves),
compared to the best-case ATLAS projection (dotted curves), for pair
production of LLPs in exotic
 Higgs decays $h \to X X$, from \cite{Chou:2016lxi}.   The three
 curves
 of each set correspond to three different  values of the LLP
 mass. 
Sensitivity up to the BBN lifetime limit~\cite{Fradette:2017sdd} is possible. }
\label{f.mathuslareach}
\end{figure}

\section{Design of the MATHUSLA detector}
\label{s.mathusla}

\begin{figure}
\begin{center}
\hspace*{-1cm}
\begin{tabular}{m{8cm}cm{7cm}}
\includegraphics[width=8cm]{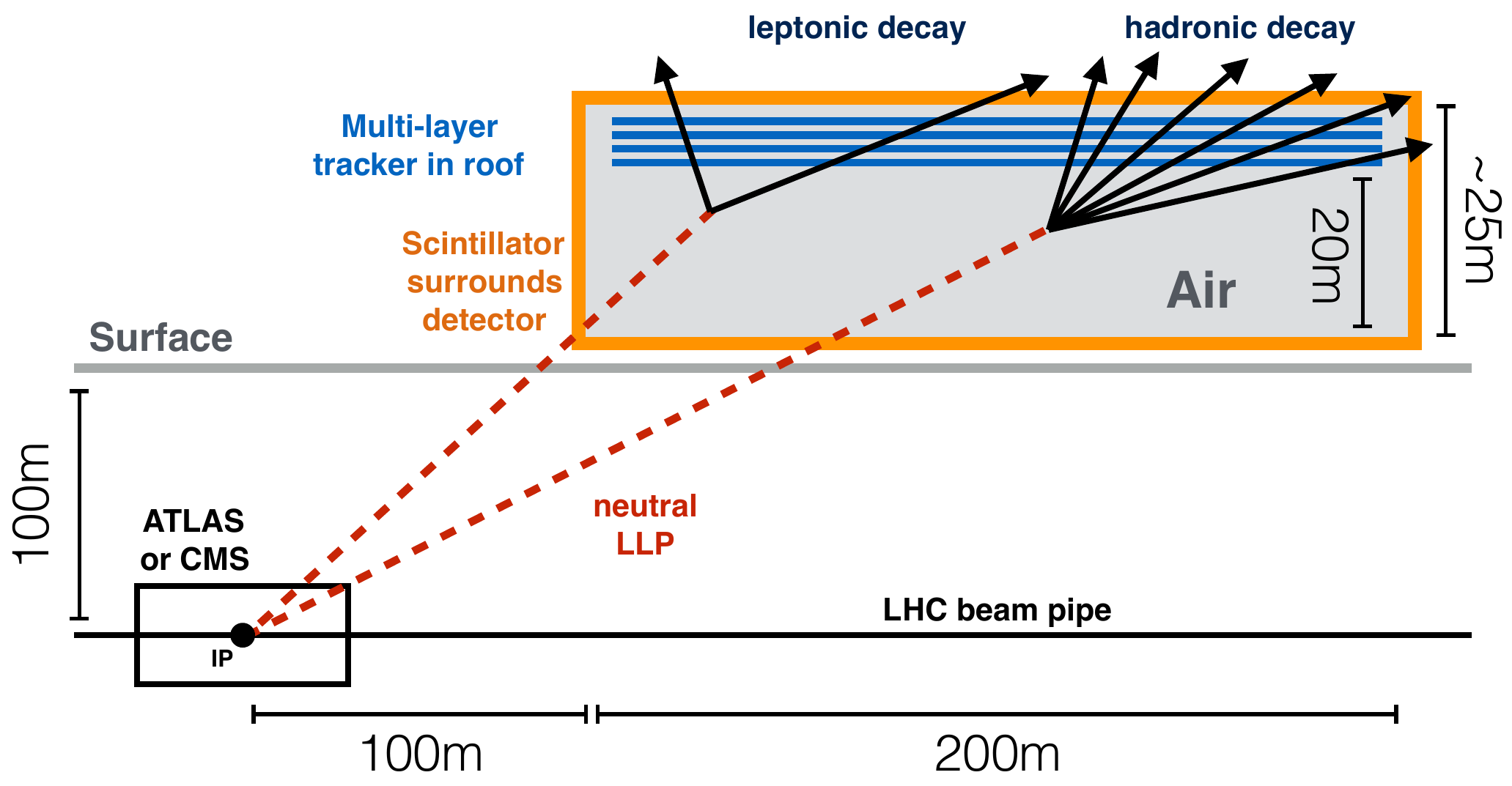}
& &
\includegraphics[width=7cm]{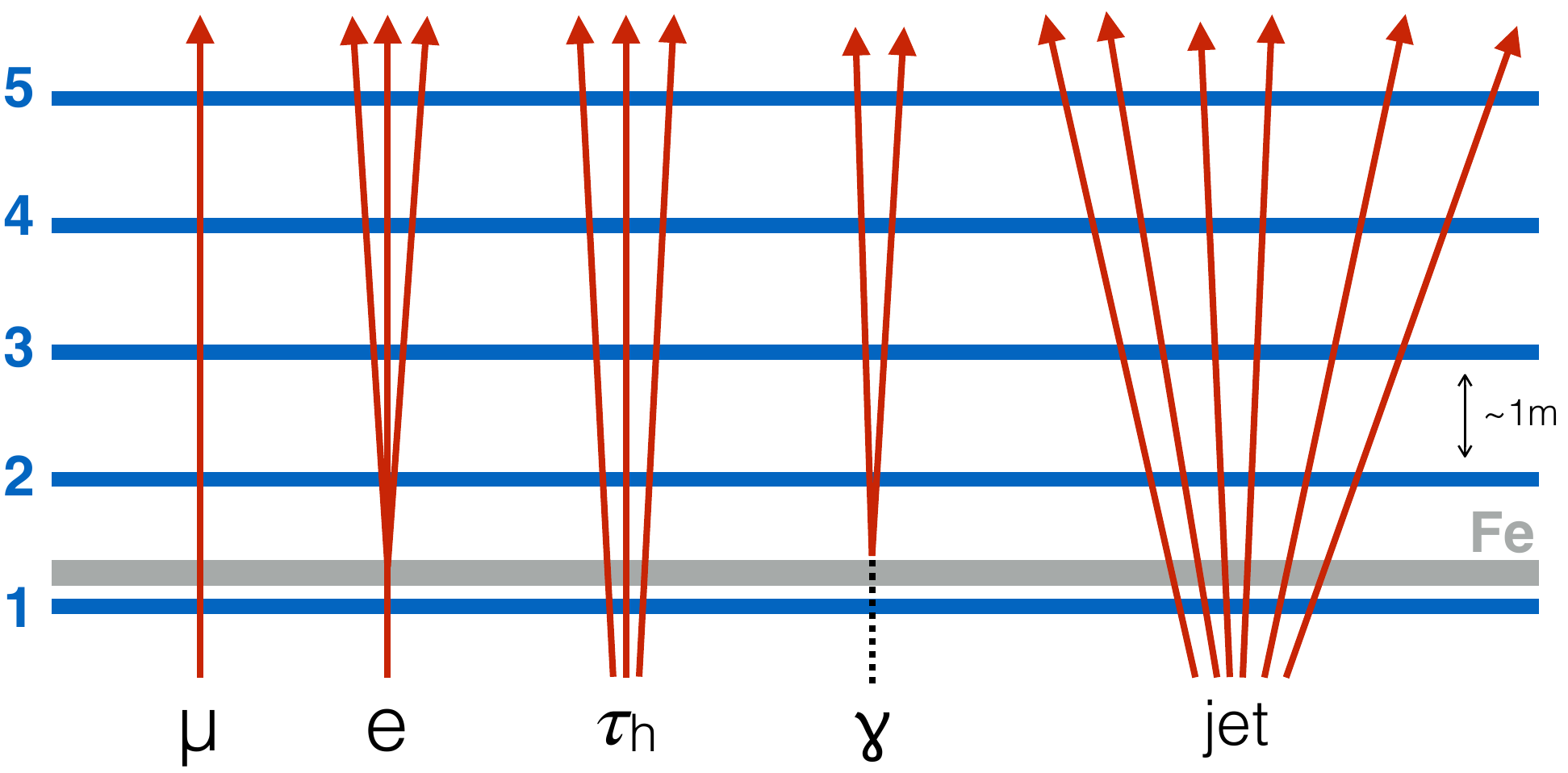}
\end{tabular}
\end{center}
\caption{
We assume the MATHUSLA detector geometry shown on the left. On the 
right, we schematically show the patterns of  charged tracks
reconstructed for 
different final states of LLP decay. }
\label{f.mathusla}
\end{figure}

For concreteness, we define a simple design for the MATHUSLA detector
that we will use in our study. This closely follows the concept
originally presented in  
\cite{Chou:2016lxi}, with one suggested modification for additional diagnostic capability.

The detector geometry relative to the LHC interaction point is shown
in 
\fref{mathusla}. MATHUSLA is an empty building of
area 40,000 m$^{2}$ and height $\sim 25$~m.   The floor, ceiling and
walls contain a  layer of
scintillator to provide a veto for charged particles emerging from
below. The veto is 
important in dealing with backgrounds, which, as is shown
in~\cite{Chou:2016lxi},
 can be reduced to negligible levels. However, it plays no role in the
 analysis we present here.

At a height of 20 m, we place the first of 5 RPC tracking layers.
 These  layers are spaced about 1 m apart, with the last layer
just below the roof.  The RPCs are arranged to record charged particle
hits with a pixel size of $1~\mathrm{cm}^2$ and a time resolution of  1 ns.
 This allows the angle of charged tracks to the tracking planes to be
determined with a precision 
of about 2 mrad. 

On the right side of \fref{mathusla}, we show schematically  the pattern of
charged tracks reconstructed for different final states of LLP
decay. Muons show up as single tracks, taus as one or three
collimated tracks, and jets as many tracks populating a relatively
large solid 
angle. 

Without additional material, electrons and muons may not be
distinguishable, while photons are invisible.
We therefore suggest a possible modification to the original design of 
\cite{Chou:2016lxi} by
inserting an un-instrumented steel sheet of several cm  thickness
 between the 1st and 2nd 
tracking layer. 
   This provides 1--2 $X_0$
  to convert photons and electrons, producing
 visible electromagnetic showers.   The thickness of the sheet for
 hadronic interactions is about 0.1
 $\lambda_i$.   This minimal detector does not allow measurement of
 the energy or momentum for any particles, but it does allow the
 various particle types to be distinguished qualitatively. 
 The details of this possible modification, including the exact
 thickness,
 type,  and location of material, as well as its viability in terms of 
cost and effect on tracking performance, are left for future investigation.

In our simulations, 
 we assume the following minimum detection thresholds on particle
 three-momenta to ensure the particles leave hits in all tracking
layers:  pions: 200 MeV, charged kaons: 600 MeV, muons:
 200 MeV, electrons: 1 GeV, protons: 600 GeV, photons:
 200 MeV.  We
 ignore charged pion and kaon decays within the detector volume.

 If the Higgs boson decays to a pair of LLPs with a 
branching ratio of 1\%, the ATLAS or CMS detector will produce 
 1,500,000 pairs of LLPs with the 3 ab$^{-1}$ of luminosity
 projected for the High-Luminosity LHC. Assuming the 
best case of a $\sim 100$~m  lifetime, we expect a sample
 of about 10,000  LLP decays  within the detector
 acceptance.  We will see below 
 that it is possible to draw interesting conclusions from samples with
 as
 few as 100~LPP events.

\section{Diagnosing LLPs produced in exotic Higgs decays}

In this section, we assume that LLPs are produced in pairs as the
decay products of the Higgs boson.  Our objective is to measure the LLP
mass and determine the dominant decay modes, using only geometrical charged particle
trajectories that can be measured by MATHUSLA.   We postpone to the
next section the problem of distinguishing this production mechanism
from other possibilities.

To open up the largest possible number of decay final states,
 we study LLP masses above the $\bar b b$ threshold, $m_X \in (15, 55)
 \gev$. The method is easily generalized to lower masses, though
 spatial track resolution may become more important for very light
 LLPs. We assume possible decays  $X \to e e, \mu\mu, \tau\tau, \gamma
 \gamma$ 
or $jj$. We write $\tau_{h,\ell}$ to refer to an explicitly hadronic or leptonic $\tau$.

In the simulations described below, we model
 Higgs production via gluon fusion  at the HL-LHC, with 
subsequent decay to two LLPs $X X$ and decay of one $X$ in the MATHUSLA
detector, by using the Hidden Abelian Higgs model 
\cite{Curtin:2014cca} in MadGraph~5 \cite{Alwall:2014hca}, matched up
to one extra jet, and showered in Pythia
8.162~\cite{Sjostrand:2006za,Sjostrand:2007gs}. In this model, the LLP $X$ is modeled as a
spin-0 particle, except in the case of `gauge-ordered' 2-jet decay, described below, where it is spin-1.
Only LLPs with an angle to the beam axis in
the range $ [0.3,0.8]$, corresponding approximately  to the angular coverage of
MATHUSLA,  are analyzed~\cite{tracking}. 

\subsection{Qualitative analysis}

We have already illustrated in the previous section and in
\fref{mathusla} that the various possibilities for the 2-body decay of
the $X$ can be distinguished qualitatively from the pattern of tracks
in the MATHUSLA detector.  After requiring at least two detected
charged tracks
 per displaced vertex, we can impose the following criteria to sort
 the events:
\begin{itemize}
\item two tracks: $\mu \mu$
\item two tracks, which shower after the first layer: $ee$
\item two showers but no hits in the first layer: $\gamma \gamma$
\item between 3 and 6 tracks: at least partially hadronic $\tau_h \tau_{h,\ell}$. 
\item more than 6 tracks: $jj$
\end{itemize}
 Without the material layer, photons are undetectable and electrons
 look like muons, but all of our other conclusions are unaffected.
Decays to $\tau^+\tau^-$ are recognized by the characteristic 1-prong
against 3-prong topology that appears in 26\% of  $\tau^+\tau^-$
decays.  The identification of  subdominant decays modes and the
measurement of their branching ratios depends
strongly on what the dominant decay mode might be.  In some cases,
there is an obvious analysis using the criteria above. 
   If the dominant decay mode is $b\bar b$, this generates backgrounds
   to other decay models that must be studied with care. 
 The full analysis of that problem is beyond the scope of this paper.

The category of $X$ decays to $jj$ contains a number of more specific
possibilities. Three simple
benchmark scenarios are:  (1)  decay to gluon jets, (2)
 ``gauge-ordered'' decay to $q\bar q$ jets
with democratic flavor content,
 as would be generated by the decay of a 
dark photon \cite{Curtin:2014cca}, and
(3)  ``Yukawa-ordered'' decay to  jets which are dominantly $b \bar b$,
 as would be  generated by the decay of a dark singlet scalar that
 mixes with the SM Higgs.    These possibilities cannot be
 distinguished on an event-by-event basis, but we will show below that
 they can be distinguished in samples as small as 100 events.
 
\subsection{Measurement of the LLP mass}

Now we discuss the determination of the LLP mass. It is crucial that
the decay vertex can be precisely located within the MATHUSLA decay
volume.  Since the LLP  $X$ originates from the nearby LHC collision
region, the vector from the point of origin to the decay vertex is very
well known.   This allows the velocity $\beta_X$ of the LLP to be
found from the geometry of the decay.

\begin{figure}
\begin{center}
\includegraphics[width=11cm]{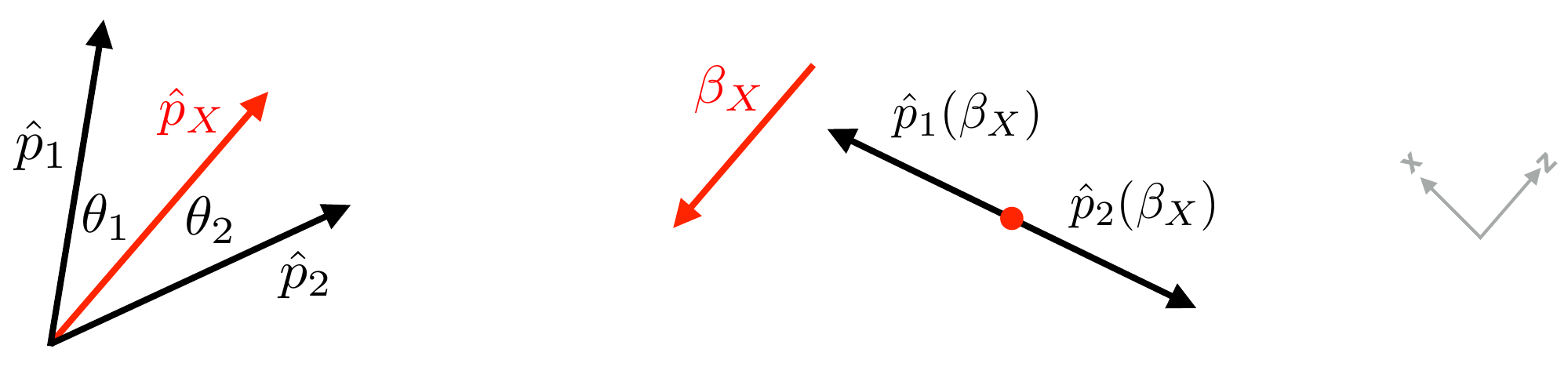}
\\
\hspace{0cm}\footnotesize Lab Frame \hspace{2cm} LLP rest frame
\end{center}
\caption{Kinematics of the two-body decay of an LLP:   The left-hand
 figure shows the angles $\theta_1$, $\theta_2$; the right-hand figure
  illustrates the boost back to the LLP rest frame in which the 2
  products are back-to-back. 
Note that $\hat p_i$ or $\hat p_i(\beta_X)$ denote momentum vectors
 normalized to unit length  \emph{in each frame}.
For convenience we work in the coordinate
  system where $\hat p_X$ is  along the $z$-axis and the decay products are in the $(x,z)$ plane (far right).}
\label{f.angles}
\end{figure}

 Consider first a decay to 2 final-state charged particles,  
such as $ee$ or $\mu\mu$. Let $\theta_1$ and $\theta_2$ be the
angles of the two decay products with respect to the $X$ direction, as
shown in \fref{angles}.  The 4-vectors of the two products then have
the form 
\beq
\label{e.pi}
     p_i = E_i (1, \pm \beta_i\sin\theta_i, 0,  \beta_i \cos\theta_i) \ \ \ \ , \ \ \ \ \ i = 1, 2
\eeqn
 with $\theta_1$ and $\theta_2$ both positive quantities and 
 $E_1\beta_1 \sin\theta_1 = E_2\beta_2 \sin\theta_2$ by
momentum balance.   Since all components are known up to a an overall prefactor, we can boost both $p_i$ back along the direction of $p_X$ until they are back-to-back, recovering the LLP rest frame. This yields
\beq
\beta_X = \frac{\beta_1 \beta_2 \sin(\theta_1 + \theta_2)}{\beta_1
  \sin{\theta_1} +
 \beta_2 \sin{\theta_2}} \ . 
\eeq{e.betaX}
 Since the distance of the LLP decay to the LHC interaction point is
much greater than the distance to the tracking planes, the precision
of the measured angles $\theta_1, \theta_2$ is simply the precision of
the measured angles between the tracks and the trackers, about 0.2\%
for  $\theta_i \sim \mathcal{O}(1)$ and approximately independent of
the uncertainty on the displaced vertex location~\cite{Knapen}.
For the two-body decays we consider,  the products will be relativistic, with
$\beta_i$ close to 1. This makes the error induced by assuming that
$\beta_i = 1$ negligible. In any case, the timing of the MATHUSLA
detector tracking elements allows each $\beta_i$ to be measured to
 5\% or better. 

\begin{figure}
\begin{center}
\includegraphics[width=10cm]{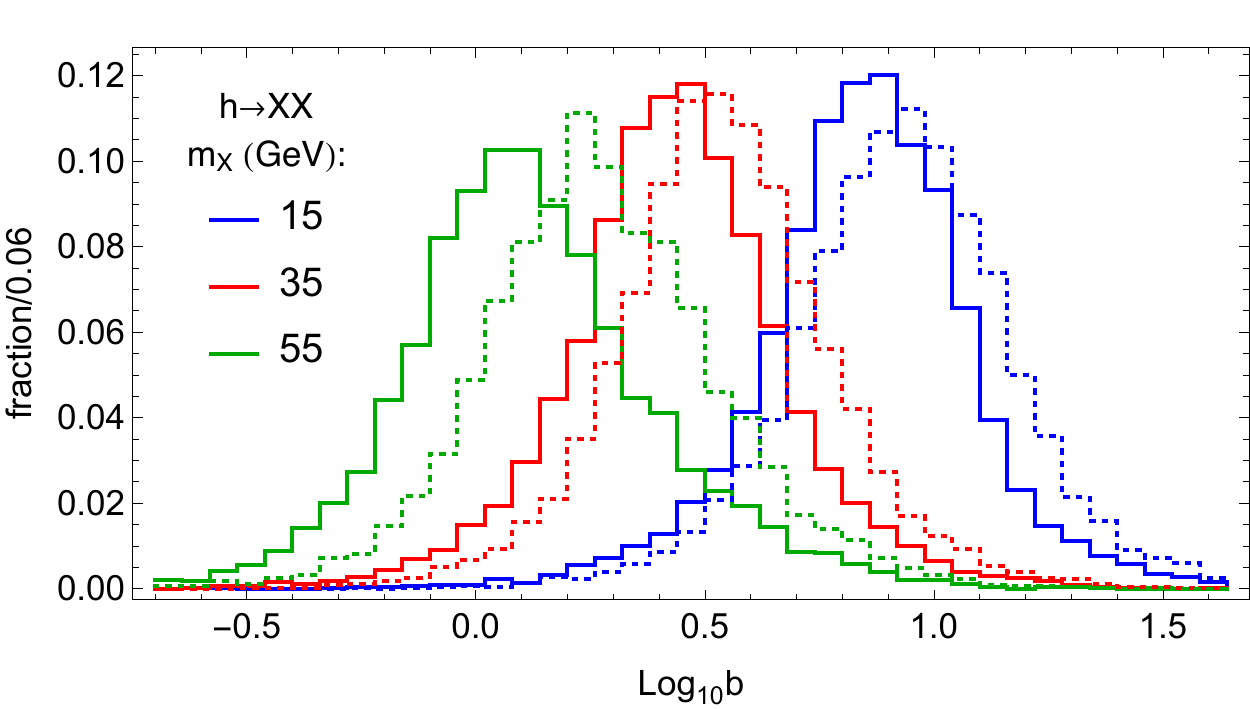}
\end{center}
\caption{
Distribution of LLP boost $b = |\vec p_X|/m$ for different LLP
masses. The solid histograms show the truth-level value of $b$, 
which is also close to the distribution of reconstructed boosts for $X$ decay to
2 charged particles.  The dotted histograms show the distribution of
reconstructed boosts
 for hadronic LLP decays using the sphericity-based method of 
\sref{Xjj} with only upwards going tracks.}
\label{f.boost}
\end{figure}

For $h\to XX$, the transverse energy of the $X$ with respect to the
LHC beam direction should be roughly $m_h/2$, so the expected mean velocity of
the produced $X$ particles decreases as the mass increases.   In
\fref{boost}, we show the expected distribution of $b = p_X/m_X$ from
our simulation for three values of the $X$ mass, illustrating this
effect.  From the figure, we estimate that a sample of 100
reconstructed events will
give the $X$ mass with   a  statistical error of about 1~GeV.  
  The
systematic error on this measurement, coming from the uncertainties in
the measurement of lepton directions, is at the part-per-mil level.

The precise knowledge of the LLP speed in each event  gives an error
on the production time 
of a few ns, making it possible to
 identify the LHC bunch crossing in which the LLP was produced.   This
 means that the event properties measured in the central detector can
 potentially be used to constrain hypotheses on the LLP production
 process.

This simple method of determining $\beta_X$ fails for hadronic LLP
decays, since jet axes cannot be reliably reconstructed from charged
particle directions alone. Fortunately, we can achieve almost
identical results using an only slightly more sophisticated method that
we  outline in \sref{Xjj}.

We now discuss this mass measurement more quantitatively for the three
cases of  $X \to \mu \mu$, $X\to jj$ and $X \to \tau \tau$.

\subsection{LLP decay to $\mu \mu$}

For $\mu \mu$ decays, \eref{betaX} gives the reconstructed LLP boost
as long as both muons hit the roof of the detector. 
This is the case in about 95\% $(50\%)$ of decays in MATHUSLA for $m_X = 15$ $(55) \gev$.
This geometrical effect is the dominant factor in the efficiency~$\epsilon$ for an event to be reconstructed by the MATHUSLA detector.

 To determine the expected precision of a mass measurement with
 $N$ reconstructed LLP decays, we conducted
 1000 pseudoexperiments
and made a maximum likelihood fit of the measured boost distribution
to 
template-functions obtained from the same boost distributions
 in the maximum-statistics limit.   For a given pseudoexperiment, we
 define   $N_{obs}$ to be the number of decays in the MATHUSLA
 detector volume and $N_{reconstructed}$ to be the number of decays in
which the tracks are oriented such that mass can be computed from the
available information.  The reconstruction efficiency $\epsilon =
N_{reconstructed}/N_{obs}$ varies from 0.95 for $m_X = 15$~GeV to 0.55
for $m_X = 55$~GeV.   The distributions of the
 reconstructed 
boosts are very close to the 
truth-level distributions shown in \fref{boost}. 
We define the expected mass
 precision,
 $\langle \Delta m / m \rangle$, to be the average spread 
of best-fit mass values amongst the 1000 pseudoexperiments.
 In \fref{Nobs}, we show the dependence of this quantity on the total
number of LLPs decaying in the MATHUSLA detector volume.  A 10\% mass measurement
 requires only 20-30 observed decays. 
 
For few reconstructed events, the precision of the mass measurement is better for heavier LLPs, while for many reconstructed events, it is better for lighter LLPs. This is not an artifact of the mass-dependent $\epsilon$  in \fref{Nobs}, but is likely due to the fact that under the assumptions of our LLP production mode in Higgs decays, the LLP mass is bounded from above. Therefore, for very few observed events, measuring a handful of very low boosts has to be due to LLPs near threshold. As the number of reconstructed events becomes large, these parameter space ``edge effects'' become less important, and the slightly narrower boost distribution of light LLPs makes their mass measurement more precise.

\begin{figure}
\begin{center}
\hspace*{-14mm}
\begin{tabular}{ccc}
\includegraphics[height=5cm]{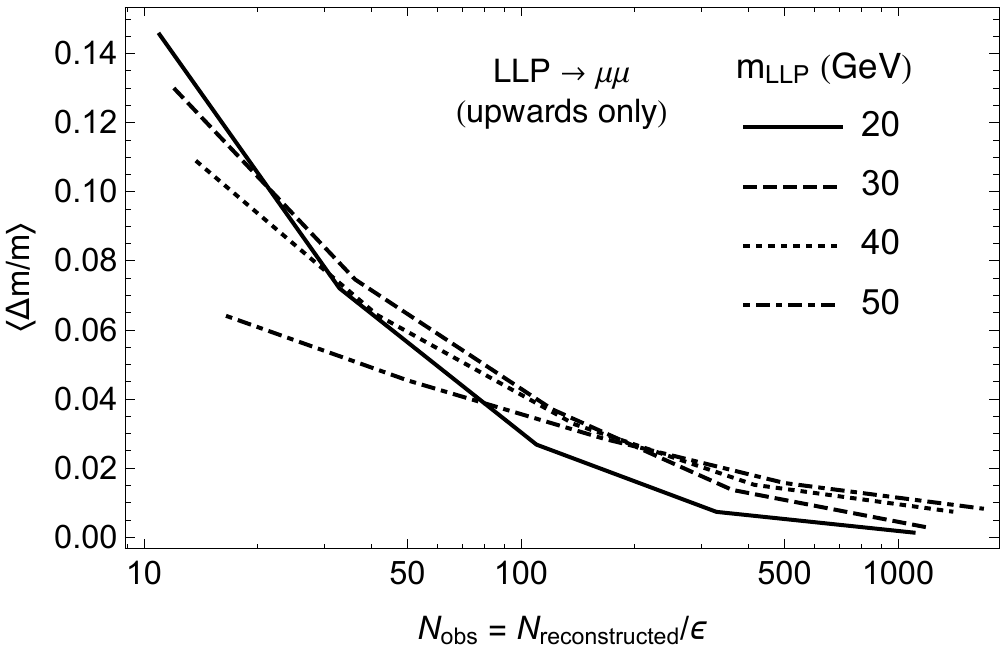}
&&
\includegraphics[height=5cm]{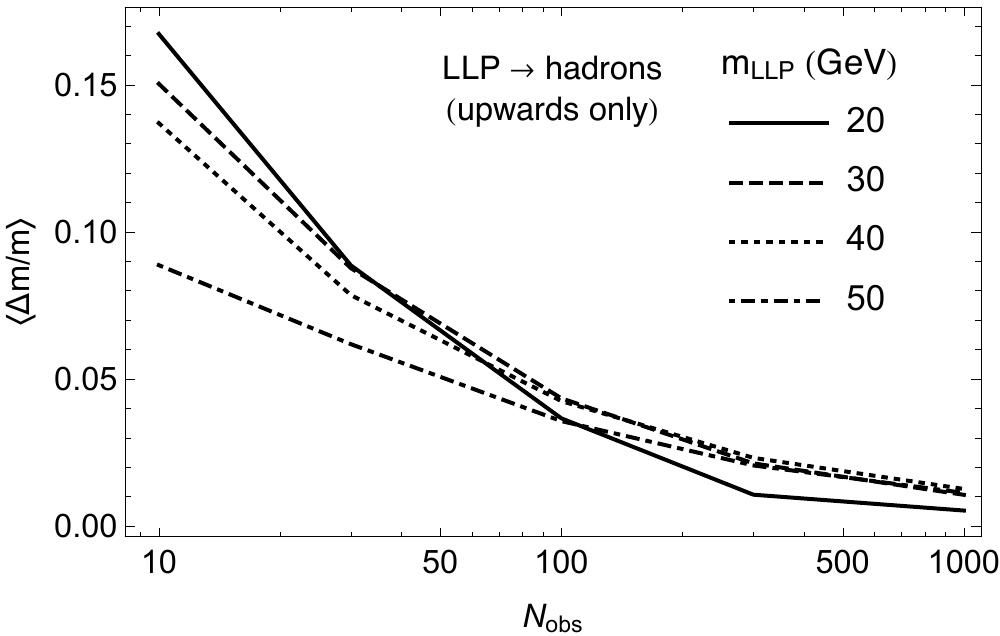}
\end{tabular}
\end{center}
\caption{
The expected number of LLP decays in MATHUSLA required to measure the LLP mass to relative precision $\Delta m/m$ using only upwards-traveling tracks, for LLP decay to muons (left) and jets (right).  
$N_\mathrm{obs}$ refers to the number of decays in the detector volume
regardless of whether the event is reconstructed. For muons, 
$N_\mathrm{reconstructed} = \epsilon N_\mathrm{obs}$ where 
$\epsilon \approx 0.95$ $(0.55)$  for 
$m_X = 15$ $(55) \gev$. For hadrons, $\epsilon \approx 1$. 
The plots show the statistical uncertainty only.  }
\label{f.Nobs}
\end{figure}

\subsection{LLP decay to jets}
\label{s.Xjj}

In our mass range of interest, LLP decays to jets produce events with
$10 - 20$ charged tracks in the detector. This is illustrated in
\fref{Ncharged} for $m_X = 15$ and $55$ GeV and our three benchmark
jet flavor compositions. This high multiplicity is a boon for several
reasons.  In terms of background rejection, a
displaced vertex with full timing information and this many
tracks is supremely difficult to fake by cosmic rays.  In terms of
signal analysis, the
multiplicity distribution  contains information about the jet
flavor composition. Furthermore, the detection efficiency for LLPs
decaying to jets (defined as the fraction of decays with at least 6
charged tracks hitting the roof) is
very close to 100\%.

\begin{figure}
\begin{center}
\begin{tabular}{ccc}
\includegraphics[width=7cm]{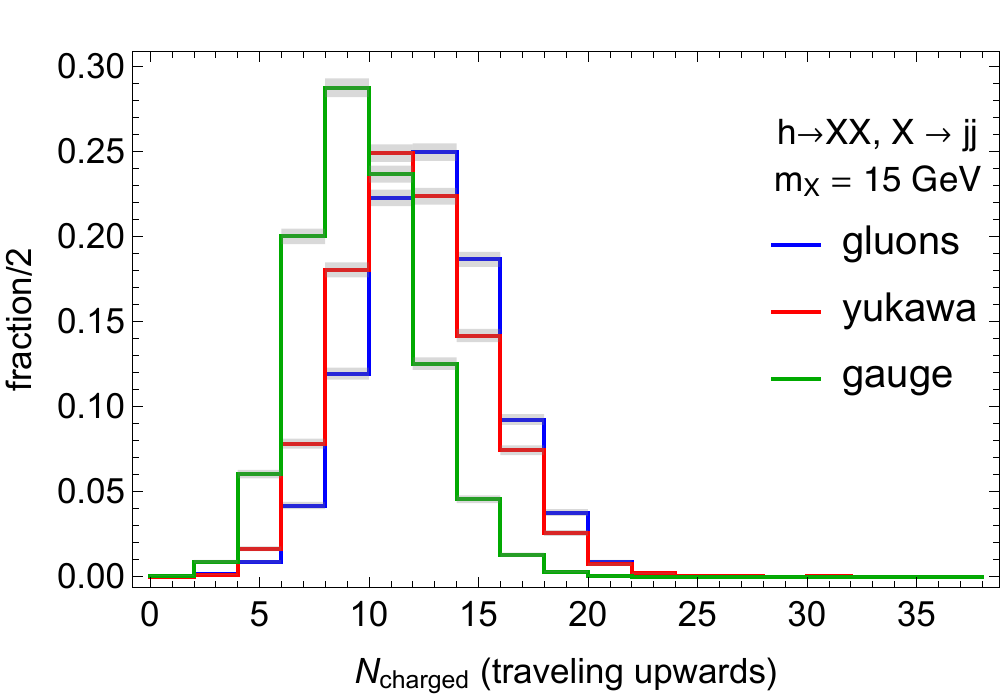}
&&
\includegraphics[width=7cm]{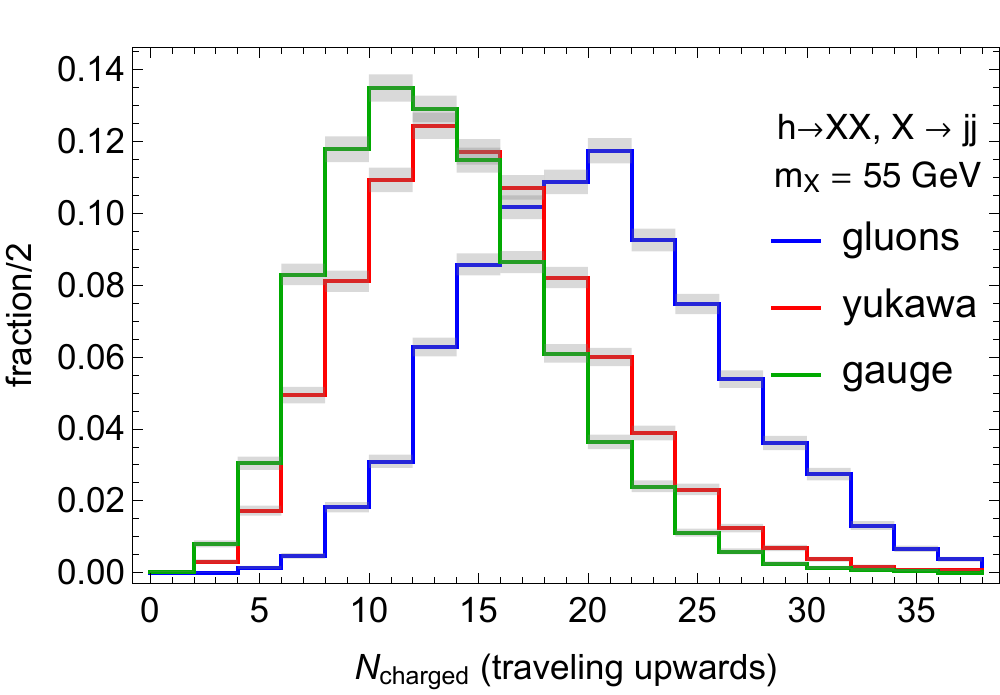}
\end{tabular}
\end{center}
\caption{
Distribution of charged particle multiplicity (only counting upwards
traveling tracks) for $m_X = 15, 55 \gev$ and gluon jets and for 
 flavor-democratic
 (gauge-ordered) and heavy-flavor dominated (yukawa-ordered)
quark  jets.}
\label{f.Ncharged}
\end{figure}

On the other hand, the high multiplicity means it is not so simple to
determine the jet directions that must be input into \eref{betaX}.
One might, for example, try to extract two jet axes
 $\hat p_a, \hat p_b$
by maximizing the quantity
\beq
V_2 = \sum_i \max(\hat p_a.\hat p_i, \hat p_b.\hat p_i)
\eeq{Vtwo}
summed over charged track momentum unit vectors $\hat p_i$ in the event.
The sum tends to be dominated by tracks carrying low fractions of the
total jet momentum.
In the absence of energy and momentum measurements, the jet axes
cannot be reliably determined by this or similar methods.

Fortunately, we can exploit the high multiplicity for a different kind
of approximate boost reconstruction.  Naively, if the LLP decays to high
multiplicity, the distribution of tracks in its rest frame should be
spherically symmetric.   Applying this assumption to individual events, we
can estimate the LLP boost event-by-event by
solving for $\beta_X$ in the constraint
\beq
\hat p_X \cdot \sum_i \hat p_i(\beta_X) = 0 \ .
\eeq{e.betasphericity}
The resulting \emph{sphericity-based boost}
distribution, using only upward going tracks and assuming all final states to be ultra-relativistic in the lab frame, is shown as the dotted
distributions in \fref{boost}.
In our simulation of LLP decays to $jj$,  this method is surprisingly
powerful.   It gives a boost distribution very close to the original boost
distribution from Monte Carlo truth.
Even more importantly, its discriminating power
 for LLP mass is almost unaffected by the deviation between these distributions.

There are several reasons why the sphericity-based method 
might give an accurate
result.  
If the parent of the LLP has spin 0 and CP violation can be ignored in its decays, the distribution of tracks in the LLP rest frame will be front-to-back symmetric on average. The same result applies if the parent has spin but is produced with zero longitudinal polarization along the LLP direction.  

For the decay to $jj$, though, a more important reason for the
accuracy of the sphericity-based method is that the sum in
\leqn{e.betasphericity} is dominated by hadrons that are soft in the
rest frame of the LLP.  The momentum distribution of these soft
particles depends only on the color flow and is
 independent  of any LLP polarization.  
Their high multiplicity ensures that the shift of any one 
sphericity-based boost measurement is much smaller than the width of
the overall boost-distribution, allowing  the LLP mass to be accurately extracted.
It should be noted that 
 these same soft hadrons are also partially responsible
 for the noticeable positive bias of the sphericity-based 
boost distribution compared to the Monte Carlo truth. 
That effect deserves a  dedicated discussion, 
which we present in the Appendix.

We can use the sphericity-based method to measure the mass of a LLP decaying
as $X \to jj$ without having to determine the jet flavor content
first.  The event-by-event precision of the sphericity-based
$\beta_X$ measurement is sufficient to determine the LHC bunch
crossing in which the LLP was created to within about 2 (6) bunch
 crossings for $m_X = 15$ (55) GeV.

We can estimate the required number of observed events $N_\mathrm{obs}$
for a given
mass measurement precision in a fashion identical to that used for $X \to \mu
\mu$ above. The only difference is the use of sphericity-based
boost distributions as the templates. The efficiency for reconstructing these events is
close to 1.  The required number of observed  events is shown
in \fref{Nobs} (right). The result is very similar to that for $X \to \mu
\mu$, with only about 20 - 30 events required for a statistical
  error on the mass measurement of 10\%.

   The mass measurement also has a systematic error from
   hadronization uncertainties in modeling the final state of the LLP
   decay. Varying Pythia tunes, we find this to be less than 1\%, but
   a full analysis should also investigate the effect of using other
   generators such as 
Herwig~\cite{Bahr:2008pv,Bellm:2015jjp} or 
Sherpa~\cite{Gleisberg:2008ta}.

The different multiplicity of charged final states can be exploited to
determine the flavor content of the $X \to jj$ final state. We make
use here of the fact that a gluon jet has a higher multiplicity than a
$b$ quark jet, which in turn has higher multiplicity than a light
quark jet.   Although the differences in the multiplicity distributions are not
large enough to identify the jet flavor on an event-by-event basis,
this becomes an effective discriminator when applied to large enough
samples.  The effect is robust even taking in account the
discrepancies in the predictions of different hadronization
schemes \cite{Gras:2017jty}.
A straightforward generalization of the mass measurement method to a 2D
likelihood fit in boost and multiplicity reveals
 that the different decay modes can be reliably distinguished with about 100 observed LLP decays.

The charged track distribution contains even more information, but it
is likely more dependent on the hadronization model and assumed
detector capabilities. For example, the minimum charged particle
velocity in each event is significantly higher for gauge-ordered jets
than for Yukawa-ordered or gluon jets for $m_X = 15 \gev$, while for
$m_X = 55 \gev$ the gluon jets have higher fraction of slower
particles than both types of quark jets. The angular
 correlations also contain information about the LLP spin.
We have not exploited this property in our analysis, but with further
study it would likely improve the diagnosis of hadronically decaying LLPs.

Finally, we point out that even though jet axes are not useful for
measuring LLP boost, a rough determination of the LLP \emph{decay
  plane} is possible
by minimizing
\begin{equation}
\sum_i (a \cdot \hat p_i)^2
\end{equation}
for choice of plane normal vector $\vec a$. The resulting decay plane
corresponds to the truth-level expectation up to a deviation angle
$\Delta \theta \sim 0.2 - 0.5$. This is crude, but it could be useful for
diagnosing significant invisible components in LLP
 decays.

\subsection{LLP decay to $\tau \tau$}

For $X \to \tau \tau$ events, each $\tau$ decay gives 1 track or 3 well-collimated tracks.  For the
events with two 1-prong decays, we use the direction of the observed track as a proxy for the
$\tau$ direction.   For events with 3-prong decays, maximization of the
quantity $V_2$ in \leqn{Vtwo} provides good approximations to the two $\tau$ directions.  Using
the two $\tau$ vectors estimated in this way, we apply the method of Section 3.2.
The fraction of events for which at least two charged
particles hit the roof of the detector is about 90\% (60\%) for $m_X = 15$ (55) GeV.
We note that the sphericity-based method described in Section 3.3 gives slightly better
results for the case of a spin 0 LLP; however, it is less robust with respect to the effects of
possible LLP polarization.  The
event-by-event precision of the
$\beta_X$ measurement is sufficient to determine the
LHC bunch crossing in which the LLP was created to within about 2 (4)
bunch
 crossings for $m_X = 15$ (55) GeV. The required number of observed
 events for a given precision of mass measurement is very similar to
 the
cases already presented in \fref{Nobs}.

\section{Determining the LLP Production Mode}

\begin{figure}
\begin{center}
\includegraphics[width=12cm]{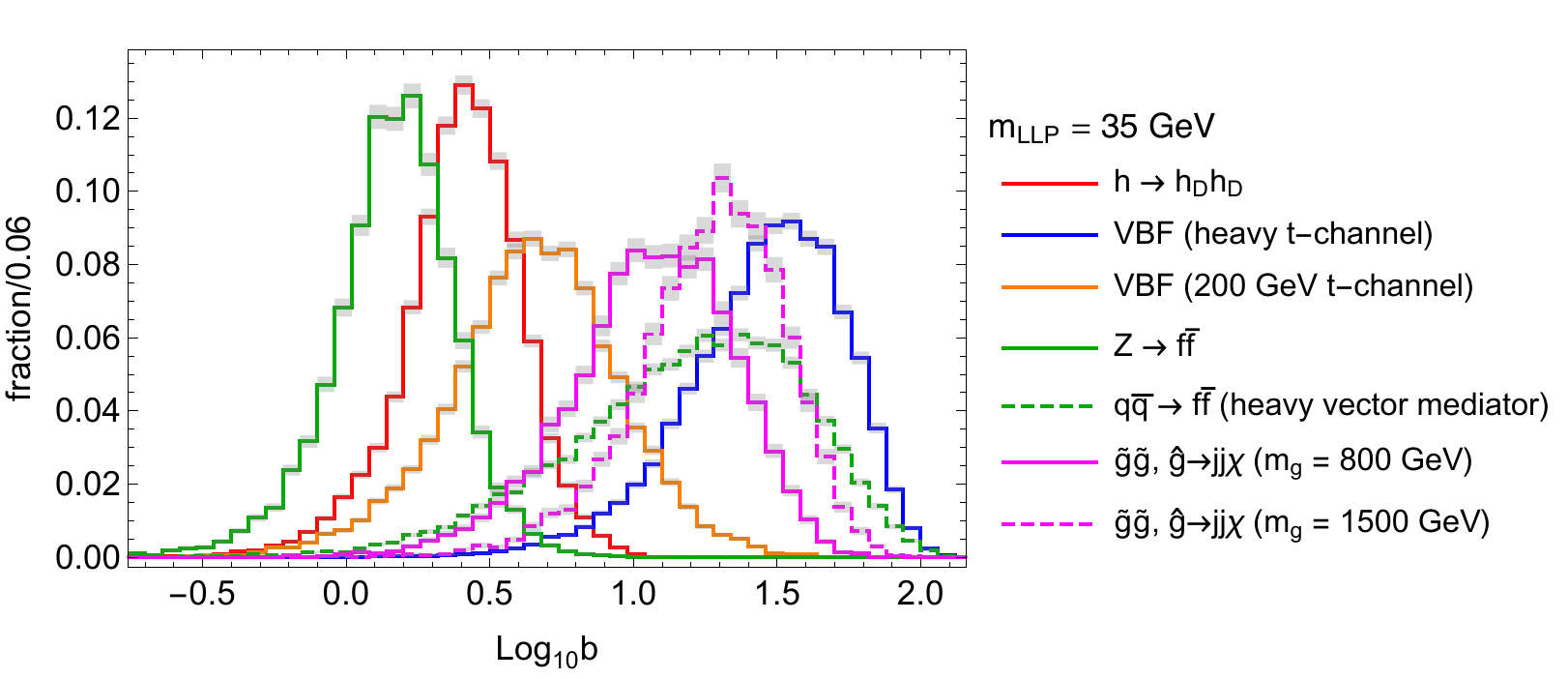}
\end{center}
\caption{
Truth-level boost distribution of a 35 GeV LLP produced in a variety
of different production modes, from left to right:  $Z\to XX$, $h\to
XX$, vector boson fusion through a 200 GeV mediator in the
$t$-channel, decay of gluinos with mass 800 and 1500 GeV, 
vector boson through through a $WW\to XX $ contact
interaction, and $q\bar q\to XX$ through a vector contact interaction.}
\label{f.alternateboost}
\end{figure}

The analysis of the previous section made explicit use of the
assumption that the LLP is 
produced in pairs in Higgs decay. However, with enough events and a library of possible production  mode hypotheses as templates, it may be possible that the LLP  production mode, decay mode and mass can all be independently determined in a global fit.  In \fref{alternateboost}, we
compare the boost distributions for a 35~GeV LLP produced through the
following mechanisms at the 14 TeV LHC:  $Z\to XX$, $h\to
XX$, vector boson fusion through a 200 GeV mediator in the
$t$-channel, 
gluino decay,
$q\bar q\to XX$ through a vector contact
interaction, and vector boson fusion  through through a $WW\to XX $ contact
interaction. 
For clarity of presentation, we have generated an
equal number of events for each sample. These six cases give six
shapes with different, distinguishable, features.

The event-by-event boost measurements also allow the LHC bunch
crossing in which the LLP was produced to be narrowed down either
uniquely  or to one of a few choices. Given the low rate per bunch
crossing of high
momentum transfer events, it is likely that, if the production event
is triggered on and recorded, it can be
identified and studied.  Even if only a small fraction of events were recorded during these
bunch crossings
this still might put interesting limits on the energy spectrum of
associated objects, distinguishing, for example, the hypotheses of  Higgs or $Z$ boson
origin from hypotheses involving $W$ fusion or contact interactions.

\section{Conclusions}

It is a real possibility that the Higgs boson decays to long-lived
particles that couple very weakly to all other particles of the
Standard Model, and that would be invisible to LHC detectors.  In
\cite{Chou:2016lxi}, a relatively simple large-volume detector was proposed to search
for such particles.  In this paper, we have explained that this
simple detector nevertheless has the power to provide qualitative and
even quantitative information about the nature of these long-lived
particles. 
This could well be our first source of information on a new sector of
particles that coexist with those of the Standard Model and open a new
dimension into the fundamental interactions.

\Acknowledgements

We thank Henry Lubatti, John Paul Chou, and the other members of the 
MATHUSLA  collaboration for helpful feedback during the completion of
this paper.
 DC also thanks Raman Sundrum and Simon Knapen for useful conversations. 
 We thank the referee for encouraging us to more closely examine the sphericity-based boost measurement, including systematic error, bias and spread. 
The work of DC  is supported by National Science Foundation grant
no. NSF-PHY-1620074 and the Maryland Center for Fundamental Physics. 
The work of MEP is supported by the US Department of
Energy under contract  DE--AC02--76SF00515.

\appendix

\section{Bias and spread of the sphericity-based boost
  measurement}

The sphericity-based boost measurement discussed in \sref{Xjj} and
shown in Fig.~\ref{f.boost} has
 three noticeable features:
\begin{enumerate}
\item The width and shape of the sphericity-based distributions are
  about the same as 
the truth-level boost distributions. 
\item For lower masses, giving a high-velocity LLP,  $\log_{10}b \gtrsim 0.5$,
  there
 is a positive bias of $\log_{10}b_\mathrm{measured} -
 \log_{10}b_\mathrm{truth} \sim 0.1$ 
which is approximately mass-independent.
\item For higher masses, giving a low-velocity LLP, the bias is
    again positive and significantly larger.
\end{enumerate}
These points are  important in preserving the
sensitivity of
 the sphericity-based boost measurement to the LLP mass. 

To investigate these effects, we found it useful to consider a toy
model of LLP decay
in which the charged final states are distributed isotropically in the
LLP rest frame
 (without respecting momentum conservation, due to the undetected
 neutral hadrons). 
The charged particle multiplicity is sampled from a Poisson
distribution  centered 
on $N_{ch} = 10$. It is instructive to consider two possibilities for the
charged final state 
momenta: either all light-like, or with mass $m = m_\pi$ and 
 energy distributed
according to a thermal 
spectrum,  $P(E) \propto \exp[-E/T]$ with $T = 140$~MeV  as a crude
model of soft 
pion emission~\cite{VanApeldoorn:1981gx}.

We used  this simple model to generate LLP decay ``events'', boosted
them to the lab frame by assuming a fixed LLP boost $b_\mathrm{LLP}$,
and 
reconstructed  the sphericity-based boosts.
For
  simplicity,
 we neglected the horizontal off-set of MATHUSLA 
from the LLP production point in this toy analysis.
For each fixed central value
$b_\mathrm{LLP}$, we defined the spread as the standard deviation of the
resulting sphericity-based boost distribution $\Delta(\log_{10}
b_\mathrm{LLP\ meas})$, and we defined the bias to be
the deviation of the average boost, 
$\log_{10} \langle b_\mathrm{LLP\ meas} \rangle - \log_{10} b_\mathrm{LLP}$.

\begin{figure}
\begin{center}
\hspace*{-5mm}
\begin{tabular}{c}
\includegraphics[width=14cm]{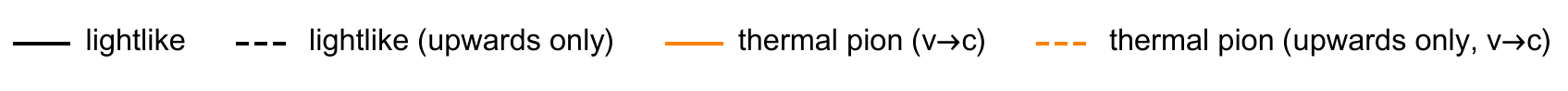}
\\
\begin{tabular}{cc}
\includegraphics[width=7cm]{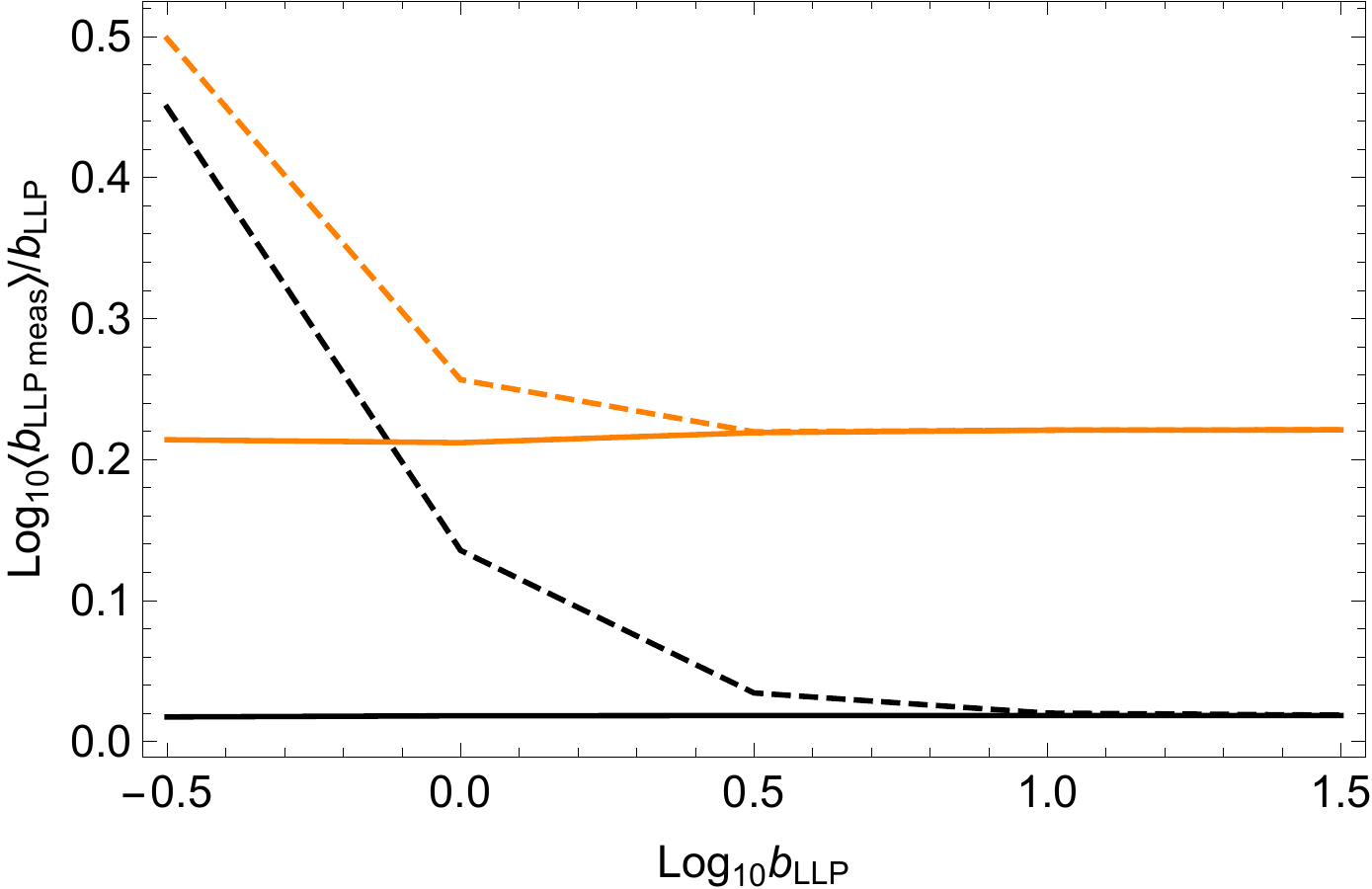}
&
\includegraphics[width=7cm]{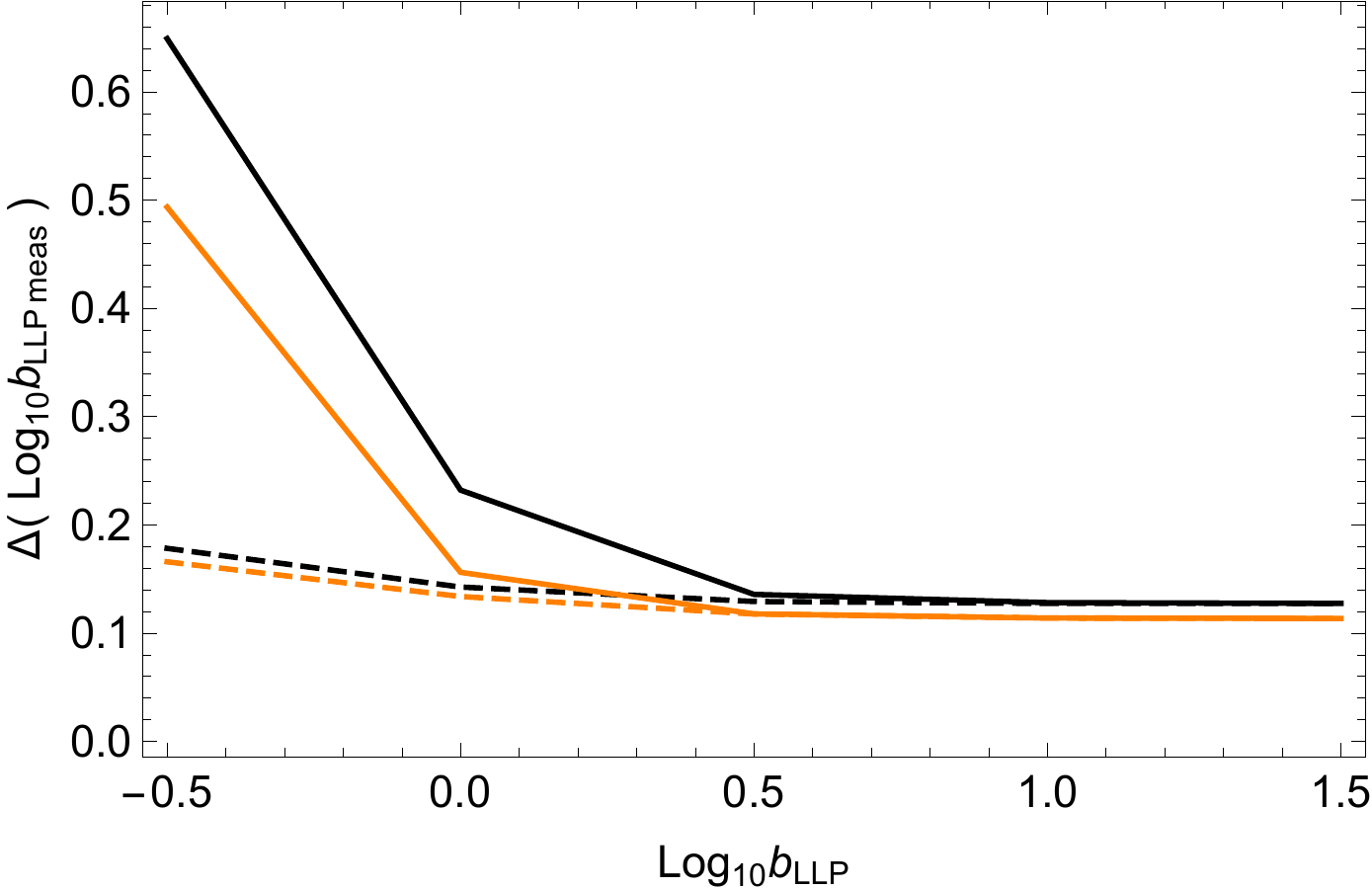}
\end{tabular}
\end{tabular}
\end{center}
\caption{
Bias (left) and spread (right) of the sphericity-based boost 
measurement for average charged particle multiplicity $N_\mathrm{ch} = 10$.}
\label{f.biasspread}
\end{figure}

The results from the toy model are shown in \fref{biasspread}. We compare
  four scenarios.  The black  lines show simulations in which all
  particles are generated with light-like momenta.   The orange lines
  show simulations in which particles are generated with a thermal
  energy distribution.   The solid lines 
  show analyses using all generated particles.   The dashed lines 
show analyses in which only upward-going  
 particles are included, as would be the case for the MATHUSLA  detector.

Consider first the values of the bias shown in the left-hand plot of
\fref{biasspread}.   The simulations with light-like momenta that
include all particles have essentially no bias.   This makes sense,
since the analysis boosts the light-like particles correctly.   The
simulations with a thermal energy distribution shows a significant
upwardd bias of 0.2, independent of velocity.    Lightlike momenta are stiffer under boosts,
and the analysis method treats these massive momenta as lightlike, so 
a larger boost is needed to balance the longitudinal momenta.  The
value of the bias  is of the same order but somewhat larger than that  in
 Fig.~\ref{f.boost}, due to the fact
that this simulation omits the leading particles in jets, which are
very relativistic.  For low LLP velocities, considering upward-going particles only 
removes a significant part of the track distribution. 
The
removal of the downward tracks 
increases the upward bias to about
0.4  in the region $b_{LLP} < 1$.

The spread, shown in the right-hand plot of \fref{biasspread}, has a
value 
of about 0.15, independent of velocity, for both
types of simulation. This is consistent with \fref{boost}.
  For analyses that consider all particles, the
spread increases at low momenta:   The true LLP  velocity is close to zero, 
so the reconstructed LLP boost is dominated by random deviations of the charged track distribution from spherical symmetry in the LLP rest frame.
For the analyses with upward-going tracks only, the reconstructed velocity is
determined by the bias, mitigating this effect.

\begin{figure}
\begin{center}
\includegraphics[width=13cm]{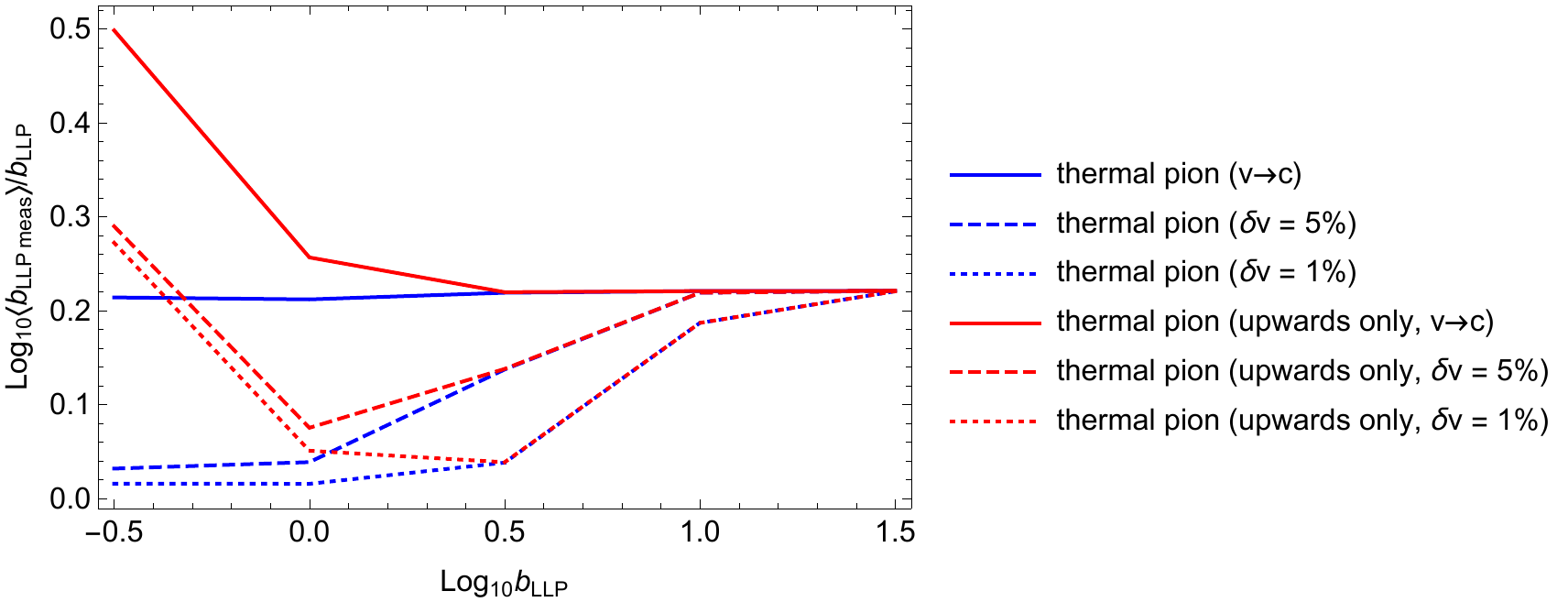}
\end{center}
\caption{
Bias of the sphericity-based boost measurement for average charged particle multiplicity $N_\mathrm{ch} = 10$, showing effect of speed measurement with precision $\delta v$ in the thermal pion case.
}
\label{f.biasthermalpion}
\end{figure}

The MATHUSLA detector has the capability of measuring the velocities
of charged particles.   A velocity measurement of 5\% will be
straightforward.  This requires recording hits to about 1 nsec
precision
over the
typical 10 m flight path through the RPC's.   A more aggressive design
might allow a velocity measurement with 1\% error. 
 Thus, it is interesting to apply velocity
information to the measured tracks to see if the bias can be reduced.
The effect on the spread turns out to be quite small. 
The results on the bias are shown in
Fig~\ref{f.biasthermalpion}. 
The solid lines correspond to treating all tracks as light-like as in
the left-hand plot in Fig.~\ref{f.biasspread}.   The dotted lines
show the effect of using the velocity information for each track,
assuming a velocity measurement with 5\% or 1\% error, spanning the
range of capabilities estimated for MATHUSLA.  Even with 5\% errors,
there is a significant effect for low LLP velocities.   The bias
returns at very low velocities when we restrict to upward-going tracks
only.   This effect of the track velocity measurement should be
considered in more detailed design studies for MATHUSLA.

\bibliographystyle{JHEP}
%\bibliography{lifetimefrontier}

\begin{thebibliography}{10}
\bibitem{whitepaper}
A large-scale theory white paper on the MATHUSLA
   physics case, 
including the examination of many other signal models, is currently in
preparation.
\bibitem{tracking}
We also consider the effect of the spatial tracking resolution and the
momentum thresholds outlined in 
\sref{mathusla}, but unless otherwise stated, their effect
 on our results is negligible.
\end{thebibliography}

\providecommand{\href}[2]{#2}\begingroup\raggedright\endgroup

\end{document}